# Quantum Gravity and the Standard Model

Sundance O. Bilson-Thompson*
*CSSM, School of Chemistry and Physics, University of Adelaide,
Adelaide SA 5005, Australia*

Fotini Markopoulou† and Lee Smolin‡
*Perimeter Institute for Theoretical Physics,
Waterloo, Ontario N2J 2W9, Canada,
and
Department of Physics, University of Waterloo,
Waterloo, Ontario N2L 3G1, Canada*

(Dated: February 28, 2006)

We show that a class of background independent models of quantum spacetime have local excitations that can be mapped to the first generation fermions of the standard model of particle physics. These states propagate coherently as they can be shown to be noiseless subsystems of the microscopic quantum dynamics[3]. These are identified in terms of certain patterns of braiding of graphs, thus giving a quantum gravitational foundation for the topological preon model proposed in [1].

These results apply to a large class of theories in which the Hilbert space has a basis of states given by ribbon graphs embedded in a three-dimensional manifold up to diffeomorphisms, and the dynamics is given by local moves on the graphs, such as arise in the representation theory of quantum groups. For such models, matter appears to be already included in the microscopic kinematics and dynamics.

## I. INTRODUCTION

Ever since the notion that geometry is dynamical was advanced in the 19th Century and, even more since that idea was realized in general relativity, there has been a dream: to unify matter with geometry and gravity by demonstrating that matter arises from singularities or topological defects in geometry. In this paper we find that this expectation may be realized in a certain class of quantum gravity theories.

In these theories, states are labeled by diffeomorphism classes of embeddings of ribbon graphs in a background topological space and the dynamics is generated by local moves on those graphs. We show that in such theories non-geometrical conserved quantum numbers arise which label emergent degrees of freedom and which may naturally be identified with elementary particles. This is because, under the commonly used evolution moves, there are conserved excitations related to the braiding of the states. Using a recent proposal by one of us which maps precisely these excitations to quantum numbers of the standard model[1], we find that, under certain mild assumptions, these theories already contain emergent structures, having properties which can be identified with conserved charges of the standard model. The dynamics of the quantum geometry may be expected to give rise to two specific observable consequences, namely the interactions of such particle states with each other (which we would hope to be consistent with electroweak and colour interactions), and the possible "erosion" o f these particle states (leading to an inconsistency with the observed stability of the first-generation leptons and quarks). In this paper we do not claim to have shown that we correctly reproduce the interactions between particles. However we do demonstrate that the dynamics of the quantum geometry does not destabilise the emergent particle states we have identified. Further work on the stability of these states is described in [2] and the issue of interactions under these and similar rules of evolution is the subject of current work.

To obtain these results we find very helpful a new point of view about how the low energy limit of a quantum gravity theory may be expected to emerge [3]. The idea is to study the low energy limit of a background independent quantum theory of gravity by asking how the states of elementary particles remain coherent when they are continually in interaction with the quantum fluctuations of the microscopic theory. The answer is that they are protected by symmetries in the dynamics. We can then analyze the low energy physics in terms of the symmetries that control the low energy coherent quantum states rather than in terms of emergent classical geometry. In [3] it was shown that one can apply to this the technology of noiseless subsystems, or NS, from quantum information theory[4]. In this framework, subsystems which propagate coherently are identified by their transforming under symmetries that commute with the evolution. These protect the subsystems from decoherence[5].

Given a particular theory of dynamical quantum geometry, however, it is not immediately apparent whether it has any such NSs. We show that in a large class of theories, it does. These are theories in which the microscopic quantum states are defined in terms of the embedding, up to diffeomorphisms, of a framed, or ribbon, graph in a three manifold and in which the allowed evolution moves are the standard local exchange and expansion moves[6, 7]. Such ribbon graphs occur in the representation theory of quantum groups[8], and arise in topological field theories and in quantum gravity models with non-zero cosmological constant [9, 10]. The procedure described in [3] reveals emergent degrees of freedom related to the braiding of the graphs. This is the first main result of this paper.

---
*Electronic address: sbilson@physics.adelaide.edu.au
†Electronic address: fmarkopoulou@perimeterinstitute.ca
‡Electronic address: lsmolin@perimeterinstitute.ca

Our second main result is that the simplest emergent local states of such theories match, with certain mild assumptions about the dynamics, precisely the first generation quarks and leptons of the standard model. To show this we employ the results of [1] where a topological formulation of preon models[11] was given. In essence, our results amount to an embedding of the topological preon model in [1] in a class of background independent quantum theories of gravity.

Preon models, first proposed in the 1970s, are models in which quarks, leptons, and in some cases vector bosons and Higgs, are composite particles, made out of just two or three elementary preons[11]. The main challenge they faced was to provide a consistent mechanism binding preons into chiral fermions. Unlike hadrons, the light quarks and leptons are much lighter than the inverse of the largest experimentally allowed binding scale for their proposed subcomponents. This made it challenging to bind such hypothetical subcomponents by means of ordinary gauge interactions[12]. The binding mechanism proposed here operates at Planck scales, below the scales at which effective field theory would be a good description. The states are bound here, not by fields, but by quantum topology, because the configurations that we interpret as quarks and leptons are conserved under the dynamics of the quantum geometry. That is, the states are bound because there are conserved quantum numbers that measure topological properties of the states.

The theories we study here are related to Loop Quantum Gravity (LQG) and spin foam models, but differ in that the graphs which comprise the states are framed, so they are represented by the embedding of a two surface in the spatial manifold. This is known to be required when the cosmological constant is non-zero[9, 10]. They also extend the description of quantum geometry used in LQG in that it becomes optional whether or not to put labels on the graphs, and the labels may be chosen from the representation theory of an arbitrary quantum group.

When these results are applied to loop quantum gravity they answer a question which goes back to the first papers on that subject, where it was shown that the diffeomorphism invariant states were characterized by the knotting and linking of loops and graphs[13–15]. The question is this: what features of geometry do the knotting and linking measure? This has been mysterious because observables sufficient to label the degrees of freedom of quantum geometry were identified in the area and volume operators, which measure combinatorial and labeling information, but which are insensitive to the topology of the embedding. The results we describe here show that some of the information in the embedding may have nothing to do with geometry, but instead describes emergent particle states.

However, as we discuss in section VIII, the most natural interpretation of the underlying states is not a geometric one, in the sense of a quantization of general relativity, but a pre-spacetime theory. The interpretation we suggest in VIII is then in essence different from LQG even though it shares the same state space and microscopic evolution moves.

The outline of this paper is as follows. In the next section we define the class of theories we will study. Section II contains the definition of the type of theories we are considering. Section III is devoted to the derivation of the first main result, which is the existence of conserved quantum numbers, preserved by the local dynamics of the quantum geometry, associated with the braiding of edges. Section IV then presents the basics of the physical interpretation of the braiding states which arises by use of the results of [1]. In sections V we then develop some tools for characterizing the details of the conserved quantities and the action of symmetry operations, which allow us in section VI to present the details of the identification of the simplest braiding states with the fermions of the first generation of the standard model. A possible identification for the higher generations is the subject of section VII. As noted above, we do not currently propose a complete description of the dynamics of the standard model, but restrict ourselves to showing the existence of particle states which are consistent with those known from the first generation of quarks and leptons. Section VIII is devoted to the discussion of several key issues beyond the scope of this paper, including locality, the mass matrix, interactions and anomalies after which we conclude in Section IX with a list of open questions.

## II. THEORIES ON RIBBON GRAPHS

We start with a compact three manifold $\Sigma$ of genus at least 3. Given a two-dimensional surface $\mathcal{S}$ in $\Sigma$, we can regard $\mathcal{S}$ as a union of *trinions*: 2-surfaces with three distinct/disjoint regions of connection (legs, for want of a better name) to other such surfaces (Fig.1). Choosing a maximal set of non-intersecting cuts on $\mathcal{S}$ gives a particular decomposition, or *scoring*, of $\mathcal{S}$ into trinions. We shall call a scored 2-surface a *ribbon graph* $\Gamma$. It is a graph in the sense of the obvious deformation of $\mathcal{S}$ into a ribbon graph with a node for each trinion and a connecting ribbon edge or leg connecting each pair of nodes.

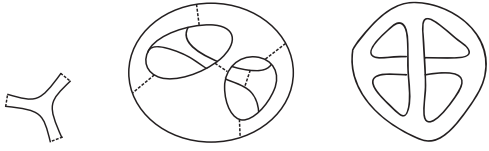

FIG. 1: A trinion and a trinion decomposition of a 2-surface $\mathcal{S}$.

A map from a ribbon graph to a quantum system can be made by associating a finite-dimensional state space $\mathcal{H}_t$ to each trinion $t$ and the tensor product operation to trinion gluings. The specific map from a trinion $t$ to its Hilbert space may depend on the choice of a quantum group $G_q$. In this case one labels each open edge of a trinion with a representation of $G_q$ and $t$ with the associated intertwiner. For the present work, we just need a functor that assigns a finite-dimensional vector space to each trinion. Thus, the state space associated to a given $\Gamma$ is

$$\mathcal{H}_\Gamma = \bigotimes_{t \in \Gamma} \mathcal{H}_t, \qquad (1)$$

with the label $t$ running over all trinions in $\Gamma$ and a sum over labels (if they are present). Nothing that follows depends on the choice of $G_q$, so we make the trivial choice of no labels.

The state space of the theory is

$$\mathcal{H} = \bigoplus_\Gamma \mathcal{H}_\Gamma, \qquad (2)$$

where the sum is over all topologically distinct embeddings of all such surfaces in $\Sigma$ with the natural inner product $\langle \Gamma | \Gamma' \rangle = \delta_{\Gamma \Gamma'}$.

Some specific examples of theories with such state spaces are: graphs labeled by quantum groups (where representations of the quantum group $G_q$ label the edges and $\mathcal{H}_t$ is the space of $G_q$-invariant tensors on $\mathcal{S}$), topological field theory, Loop Quantum Gravity in both $2+1$ and $3+1$ (in which case the algebra is $SU_q(2)$ and the quantum deformation is defined in terms of the cosmological constant $\Lambda$ by $q = \exp\{\frac{2\pi}{k+2}\}$, with $k = \frac{6\pi}{G\Lambda}$ [10]) and, in general, the functorial construction of topological quantum field theories when the source category is 2-dimensional surfaces $\mathcal{S}$.

Local dynamics on $\mathcal{H}$ can be defined by excising subgraphs of $\Gamma$ and replacing them with new ones. The generators of such dynamics are given graphically in Fig.2. Given a ribbon graph $\Gamma$, application of $A_i$ results in

$$\hat{A}_i |\Gamma\rangle = \sum_\alpha |\Gamma'_{\alpha i}\rangle \qquad (3)$$

where $\Gamma'_{\alpha i}$ are all the ribbon graphs obtained from $\Gamma$ by an application of one move of type $i$ ($i = 1, 2, 3$).

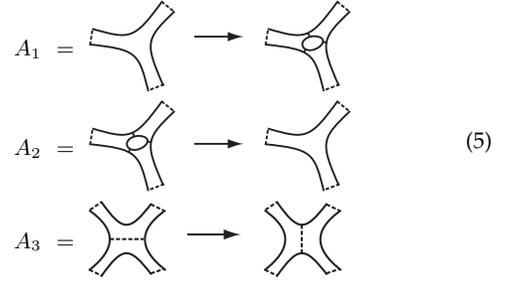

FIG. 2: The three generators of evolution on the ribbon graph space $\mathcal{H}$. They are called expansion, contraction and exchange moves.

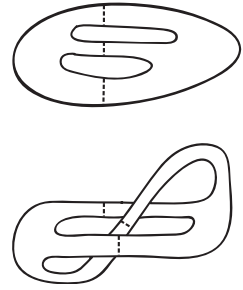

FIG. 3: These two graphs are not distinguished by the usual trinion decomposition of $\mathcal{H}$.

Together with the identity, these moves generate the *evolution algebra*

$$\mathcal{A}_{\text{evol}} = \{\mathbf{1}, A_i\}, \qquad i = 1, 2, 3 \qquad (4)$$

on $\mathcal{H}$.

This is a basic description of such theories which will suffice for the purposes of this paper.

## III. CONSERVED DEGREES OF FREEDOM

Note that the decomposition (2) of the state space $\mathcal{H}$ sums over topologically distinct embeddings of the surfaces in $\Sigma$. For example, the two graphs of Fig.3 may correspond to the same state in $\mathcal{H}_\Gamma$ but correspond to distinct states in $\mathcal{H}$. They are distinct physical states because they are braided differently.

The basic point of this paper is a simple one and can be summarized as follows. The braiding content of states in $\mathcal{H}$ remains invariant under the action of $\mathcal{A}_{\text{evol}}$. Thus, any physical information encoded in the braiding will be conserved under this evolution. As the quantum geometry evolves the conserved structure represented by the braiding will not necessarily remain localized with respect to the graph. In general as a graph evolves a configuration that appears initially to be a localized braid becomes "spread out". Nevertheless there is a conserved quantity associated with the braiding. This is to be expected and is discussed in detail in the penultimate section.

Let us first state the precise sense in which the braiding is conserved. It is easiest to do so using the notion of a noiseless subsystem from quantum information theory [4]. Given a finite-dimensional state space

$$\mathcal{H} = \mathcal{H}_A \otimes \mathcal{H}_B \oplus \mathcal{K}, \qquad (6)$$

where $\mathcal{K}$ is the orthogonal complement of $\mathcal{H}_A \otimes \mathcal{H}_B$ in $\mathcal{H}$, and a completely positive map $E$ on $\mathcal{H}$, the subsystem $B$ is *noiseless* under $E$ if, for all $\rho^A \in \mathcal{H}_A$, $\rho^B \in \mathcal{H}_B$, there exists $\sigma^A \in \mathcal{H}^A$ such that

$$E\left(\rho^A \otimes \rho^B\right) = \sigma^A \otimes \rho^B. \qquad (7)$$

That is, $\mathcal{H}_B$ viewed as a subsystem of $\mathcal{H}$ is conserved under $E$ and any information encoded in $\mathcal{H}_B$ will be protected under $E$.

Given $\mathcal{H}$ and $E$, it is a non-trivial exercise to find the decomposition (6) that reveals the noiseless subsystem. Some of the literature devoted to this task can be found in [4]. In our case, however, it is easy to find a noiseless subsystem in the state space we are considering, when $E$ is in $\mathcal{A}_{\text{evol}}$. Let us consider a given $\Gamma$ shown in Fig.4. The surface $\Gamma$ consists of two nodes that share three edges. These edges may be braided and twisted an infinite number of ways. For each choice of braiding and twisting there is a basis element in $\mathcal{H}$ and these span subspace of $\mathcal{H}$ we will call $\mathcal{H}_{\text{braid}}$. This can be decomposed as

$$\mathcal{H}_{\text{braid}} = \mathcal{H}_\Gamma^T \otimes \mathcal{H}_\Gamma^B \qquad (8)$$

where $\mathcal{H}_\Gamma^T = \sum_{t \in \Gamma} \mathcal{H}_t$ contains all trinions (unbraided and untwisted) in $\Gamma$ and $\mathcal{H}_\Gamma^B = \bigotimes_{t \in \Gamma} \mathcal{H}_b$ where $\mathcal{H}_b$ are state spaces associated with braids and twists in between the trinions (as illustrated in Fig.4).

We do not, at this stage, need to be explicit about the different kinds of braids that appear in the second factor. Our task will be to find out what the physical content of the $\mathcal{H}_\Gamma^B$ may be.

With the new decomposition, it is straightforward to check that operators in $\mathcal{A}_{\text{evol}}$ can only affect the $\mathcal{H}_\Gamma^T$ and that $\mathcal{H}_\Gamma^B$ *is noiseless under* $\mathcal{A}_{\text{evol}}$. This can be checked explicitly by showing that the actions of braiding and twisting of the edges of the graph and the evolution moves commute.

Note that this decomposition also makes explicit the fact that, starting with a graph $\Gamma$ and applying moves in $\mathcal{A}_{\text{evol}}$ on it, one cannot reach any other graph but only graphs with the same braiding content as $\Gamma$.

## IV. THE PHYSICAL CONTENT OF $\mathcal{H}^B$

We have seen that the braiding and twisting of the ribbon graphs is conserved under evolution. As was pointed out in [3], a noiseless subsystem (or nearly noiseless) in a background independent theory means that there are degrees of freedom that are noiseless under the Planckian evolution $\mathcal{A}_{\text{evol}}$ and therefore should appear in the low energy theory. The question is of course what is the physical content of the braids and the twists.

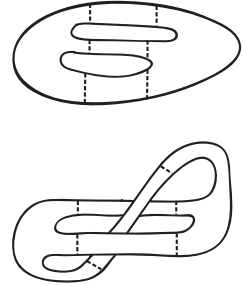

FIG. 4: The new decomposition is in terms of untwisted and unbraided trinions and braids connecting them.

It is remarkable that the noiseless subsystem we uncovered fits very naturally with the recent proposal by one of us of a preon model for the first generation of the standard model based on ribbon braids[1]. The basic idea of that proposal is to give the following physical interpretation to the braids:

- Twist is interpreted as $U(1)$ charge, so that a $\pm 2\pi$ twist in a ribbon represents charge $\pm e/3$.

- The simplest non-trivial braids can be made with three ribbons and two crossings, as in Figure 5. It is remarkable that with a single condition, these map to the first generation of the standard model [1]. This is seen in Figures 16-19 and will be recalled in more detail below.

In the rest of this paper we explore the embedding of the model of [1] into quantum geometry. We will need a more detailed description of the conserved quantities, which we will develop in the next section. For example to understand the identification of twist with charge we have to understand how charge conjugation symmetry acts on braids.

## V. INVARIANTS OF EMBEDDINGS, SYMMETRIES AND CONSERVED QUANTUM NUMBERS.

We saw in Section III that braiding and twisting are preserved under the evolution moves, giving rise to the emergence of the topological preon model from models of quantum gravity. To ensure that the precise topologies used in the topological preon model of [1] are distinct states of quantum gravity it is very helpful to work with explicit topological invariants which can easily be proved to be invariant under the evolution moves shown in Figure 2. This is the task of the present section. We will need the reduced link invariant of a ribbon graph embedding and the discrete symmetries on the braids which we define next.

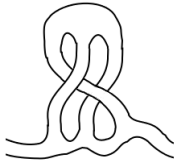

FIG. 5: A simple braid inside a ribbon graph

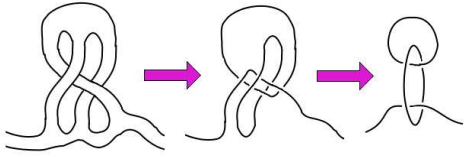

FIG. 6: The reduced link of the braid in Figure 5.

### A. The reduced link invariant

Let $\mathcal{L}[\Gamma]$ denote the the *link* of a ribbon graph embedding. This is the set of curves in $\Sigma$ obtained by considering each edge of the ribbon graph as a curve in $\Sigma$. We define the *reduced link* of a ribbon graph embedding, denoted $\mathcal{RL}[\Gamma]$, to be $\mathcal{L}[\Gamma]$ minus any unlinked unknotted closed curves.

As an example, consider the simple piece of a ribbon graph shown in Figure 5. Its reduced link is shown in Figure 6. We see it evolved in Figures 7 and 8. It is easy to verify that the reduced link is unchanged. We emphasize that, although the state does change (because two states that differ by a local move are orthogonal under the inner product), the reduced link does not.

Figures 10 to 12 further illustrate the construction of the reduced link of a ribbon graph. The reduced link can be simplified by using the standard Reidemeister moves.

It is easy to see that both the link, and the reduced link of a ribbon graph are unchanged by the action of an exchange move. While an expansion move does change the link of a ribbon graph, it does not change the reduced link. An example of this can be seen by comparing Figure 10 and Figure 12. We thus see that the reduced linking number $\mathcal{RL}[\Gamma]$ is an invariant, preserved under the local moves.

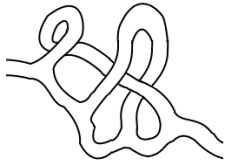

FIG. 7: A possible evolution of the braid in Figure 5 under an exchange move.

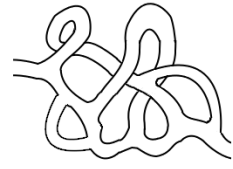

FIG. 8: Further evolving the braid in Figure 7 by expansion moves.

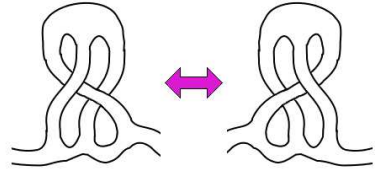

FIG. 9: The effect of a parity transformation on a braid.

We can define operators that represent the reduced link invariants. Let $K$ be a link in $\Sigma$ and let $J(K)$ be any complex valued link invariant. Given a ribbon graph $\Gamma$, let $\mathcal{RL}(\Gamma)$ be its reduced link invariant. Then define

$$J(\Gamma) = J(\mathcal{RL}(\Gamma)). \tag{9}$$

Then we can define an operator $\widehat{J}$ by its action on a basis

$$\widehat{J}|\Gamma\rangle = J(\Gamma)|\Gamma\rangle. \tag{10}$$

From the fact that the reduced link invariant is preserved by the evolution moves, we deduce that $\widehat{J}$ is in the commutant of $\mathcal{A}_{\text{evol}}$. This result tells us that there are conserved quantum numbers emergent from dynamical quantum geometries.

### B. Subsystems of ribbon graphs

There are cases where the reduced link of a portion of the graph gives a closed link, unlinked and unconnected to the rest of the graph. This is illustrated in Figures 10 to 12. Since topological invariants of the reduced link are preserved under evolution, this feature is conserved. This gives us an invariant definition of a subsystem. We can then consider the reduced link of such a subsystem.

We can define,

$$\mathcal{RL}_S(\Gamma) = \mathcal{RL}(\Gamma_S). \tag{11}$$

By the same logic as before, this is preserved under evolution moves. Hence given any link invariant $J$ we have again an operator

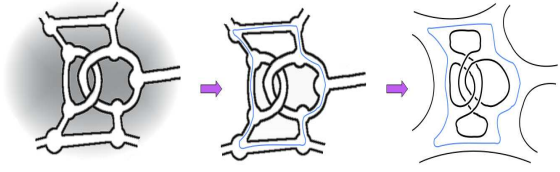

FIG. 10: Constructing the link of a ribbon graph. The blue line is a boundary which separates the region of interest from the rest of the ribbon graph.

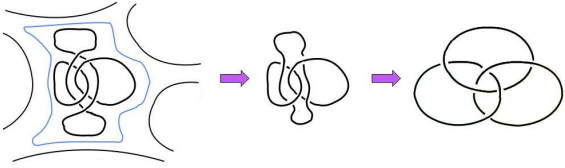

FIG. 11: The reduced link, taken by excluding unlinked, unknotted curves from the link obtained previously.

$$\widehat{J}_S|\Gamma\rangle = \sum_S \mathcal{RL}_S(\Gamma)|\Gamma\rangle. \tag{12}$$

where the sum is over disconnected pieces of the reduced link of the graph. The operator applies the link invariant to each of the disconnected pieces and sums them. In the case there is a boundary we can distinguish the components connected to the boundary, which would not be included in the sum.

This implies that given the consistent splitting, $\widehat{J}_S$ also is in the commutant of $\mathcal{A}_{\text{evol}}$. This means that we have conserved quantities connected with subsystems. This is true in spite of the fact that there are evolution moves which generate interactions between the subsystems and the rest of the ribbon graph.

We may note that the subsystems cannot in general be considered "localized" as they may become arbitrarily large and complex as the quantum geometry evolves. But nevertheless, there is a notion of subsystem and it will be useful to give some simple examples. Let us then consider a system defined by a simple set of local configurations of the form of Figure 13. They are a braided set of $n$ edges, which are joined on both ends by a set of connected nodes (or just one node.) At least one of the sets of nodes is connected to the rest of the ribbon graph. We will call them encapsulated braids.

It is clear that the form of Figure 13 is preserved under the evolution moves. Local invariants of the braidings then label persistent structures of the kind discussed in [3]. The commutant of the interaction algebra will then include elements of the braid group that connect braids with distinct reduced link invariant.

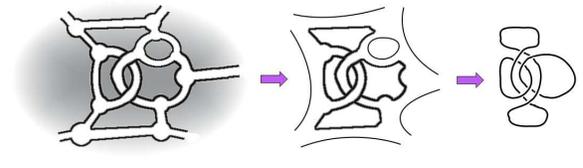

FIG. 12: The reduced link of a ribbon graph is unaffected by expansion moves.

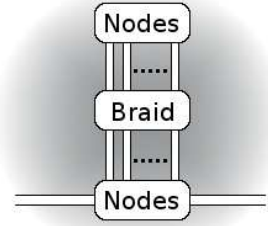

FIG. 13: Local structures, consisting of a braid joined on each end by nodes. One or both sets of nodes are connected to the rest of the ribbon graph.

### C. Discrete symmetries of braids

It will be useful in classifying states to introduce the discrete symmetries which can act on subsystems of the kind just described.

*Parity.* Let $\mathcal{T}[\Gamma]$ be any projection of the embedding of the graph onto a two-dimensional plane. Let $\mathcal{PT}[\Gamma]$ be its parity inversion, shown in Figure (9), which for a braid is equivalent to a left-right inversion, while not affecting the handedness of any twists on the strands. (Note that the definition of the parity transformation is independent of the two-dimensional plane used for the projection.) This is the projection of a graph $\mathcal{P}[\Gamma]$ which is the parity inversion of $\Gamma$. If $\mathcal{P}[\Gamma]$ is diffeomorphic to $\Gamma$ we say that the chirality of $\Gamma$ is even. Otherwise the chirality is odd, because it will be true that $\mathcal{P}^2[\Gamma] \sim \Gamma$. We then arbitrarily label the chirality of either $\Gamma$ or $\mathcal{P}[\Gamma]$ to be left-handed, so the other is right-handed. It is easy to show that the chirality of a ribbon graph embedding is invariant under both exchange and expansion moves.

*Charge Conjugation.* To define charge conjugation precisely it is helpful to recall the braid algebra, which are the discrete operations that generate braids[8]. Braids (as in Figure 13) are generated by a finite set of braiding and twisting moves, displayed in Figure 14. (Note that while braiding occurs in ordinary graphs, twisting requires framed graphs.) We can call a braid move (either a braiding or a twisting) generically $m$. A generic braid on $n$ strands generated by $p$ braid moves is then of the form of $\mathcal{B} = m_p....m_2m_1$. The possible braid moves form an algebra, which is described in [8]. The identity in the algebra $\mathbf{1}_n$ consists of $n$ unbraided and untwisted

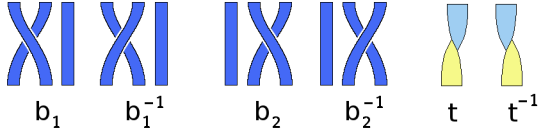

$b_1 \quad b_1^{-1} \quad b_2 \quad b_2^{-1} \quad t \quad t^{-1}$

FIG. 14: The basic braiding moves on three strands, and the basic twisting moves.

strands.

Let us fix $n$ and the sets of nodes at the two ends. Then the possible braids that can come in the box define a Hilbert space, $\mathcal{H}_{n\text{braids}}$ within which the possible braids $\mathcal{B}$ give a basis $|\mathcal{B}\rangle$. The generators $b_i$ and $t_i$ are represented by unitary operators in this space [22]. There is a natural vacuum $|0\rangle = |\mathbf{1}_n\rangle$ which is $n$ unbraided strands. Any braid is then defined by a sequence of braid moves acting on $|0\rangle$,

$$|\mathcal{B}\rangle = m_p....m_2 m_1 |0\rangle. \qquad (13)$$

For each choice of $n$ and end nodes, $\mathcal{H}_{n\text{braids}}$ defines a possible system's Hilbert space to describe emergent particle states.

We can now describe how the discrete symmetries act on the braids. We have already defined the parity operator $\mathcal{P}$. This easily restricts to an operator on $\mathcal{H}_{n\text{braids}}$. There is also a natural charge conjugation operator, $\mathcal{C}$, defined by

$$\mathcal{C} \circ \mathcal{B} = \mathcal{B}^{-1} = m_1^{-1} m_2^{-1} ... m_p^{-1} \qquad (14)$$

to be the braid constructed by making the inverses of each move, in the reverse order. It is straightforward to show that $\mathcal{P}$ and $\mathcal{C}$ commute with the evolution rules, and are hence observables.

It follows from (14) that

$$(\mathcal{C} \circ \mathcal{B})\mathcal{B} = \mathbf{1}. \qquad (15)$$

This suggests the identification of twist with an abelian charge, and $\mathcal{C}$ with charge conjugation. As noted above, we in fact identify twist with U(1) charge.

## VI. IDENTIFICATION OF THE FIRST GENERATION FERMIONS OF THE STANDARD MODEL

We have now all the structure we need to present the second main result, which is the embedding of the topological preon model of [1] in the class of quantum gravity theories we defined above.

To do so we need to make the following additional assumptions, which are common also to preon models, including that in [1].

- The lightest states are the simplest non-trivial braids made of ribbons with no twists (no charge) or one full twist (positive or negative charge).

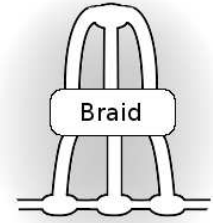

FIG. 15: The simplest three strand braids. The ovals signify nodes.

- Quantum numbers are assigned only to braids with no positive and negative charge mixing. Such a rule is necessary in preon models as discussed[1].

Ultimately such rules have to arise from the dynamics. However to study them we need to have access to the effective dynamics so that we can study the mass matrix of the theory. In Section VIII, we discuss this further; for the present, these remain ad hoc assumptions.

We will consider braids of the form of Figure 13. For each number of strands $n \geq 3$ there are an infinite number of states in the corresponding state space, $\mathcal{H}_{n\text{braids}}$. These can be ordered in terms of minimal numbers of crossings required to create the braid. A common way to understand this is to represent a graph by a projection $\mathcal{T}[\Gamma]$. For every ribbon graph embedding $\Gamma$ there is a minimal number of crossings, in any such projection, $m[\Gamma]$.

Our first assumption above amounts to considering minimal crossing number $m[\Gamma]$ to be a measure of the complexity of a graph so that the braids with the lowest crossing number will become the first generation. We then analyze the structure of the braids with the lowest crossing numbers.

For three strands there are non-trivial braids with two crossings. The simplest structure these can have is a single node at the top tying the strands together, while the strands join to the environment through the bottom as shown in Figure 15.

We therefore begin by analyzing these. Let us first neglect twists. Assuming that the twists on all generators are zero, the possible non-trivial topologies are shown in Figure 16. We see that two possible mirror-image states can be formed, which we may call left-handed and right-handed states. The remarkable fact that the simplest braids are chiral will allow us to propose the identification of them with chiral fermion states.

We now allow any or all of the strands to carry a single twist, with either positive or negative orientation. This gives rise to $2 \times 3^3 = 54$ states.

Let us call this space of states $\mathcal{V}_3$ for the states of three strand braids. It is the sum of the left- and right-braided

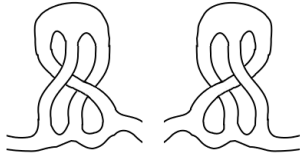

FIG. 16: The electron neutrino and anti-neutrino – two uncharged states.

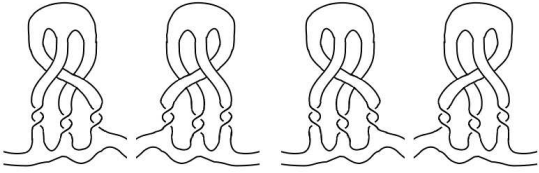

FIG. 17: The electron and positron — four maximally charged states.

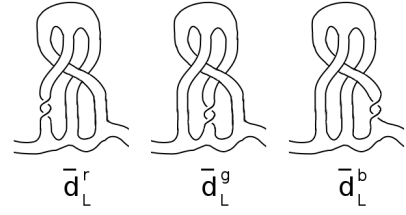

FIG. 18: The left-handed down states-showing tripling of states for fractional charge.

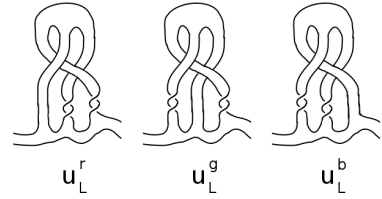

FIG. 19: The three colours of left-handed up states.

sectors:

$$\mathcal{V}_3 = \mathcal{V}_3^L \oplus \mathcal{V}_3^R \qquad (16)$$

We note that

$$\mathcal{P}: \mathcal{V}_3^L \to \mathcal{V}_3^R \qquad (17)$$

and vice versa.

However, the rules given above tell us to ignore those braids that contain both plus and minus charges (or twists). We then define the reduced subsector $\mathcal{V}_3^{red}$ which do not mix positive and negative twists, and which only have single twists. Using the results of [1] we now see that the states in the reduced subsector correspond to the 15 fermions of the first generation of the standard model of particle physics. It is straightforward to show that these 15 states have distinct reduced link invariants, hence together with parity these distinguish the 30 distinct chiral states of the first generation of the standard model. The preceeding arguments then establish that symmetries of the standard model, which transform these states among themselves, are conserved under the dynamics of the quantum gravity theory.

From [1], we can classify these states in terms of twists:

- There are four fully-charged states, that is, states with a single twist on each strand, shown in Figure 17. It then makes sense to identify twist $t$ with electric charge,

$$t = \frac{e}{3}. \qquad (18)$$

In this case we can identify the fully charged states with the electron and positron.

- There are two neutral states, one left-braided, one right-braided, which we can identify with the neutrino and antineutrino, shown in Figure 16. We note that they are both $\mathcal{C}$ and $\mathcal{P}$ conjugates of each other.

- There are the partially charged states, with one and two +s, and the rest zeroes. These are shown in Figures 18 and 19. These are the quarks, with total charges $\pm\frac{1}{3}$ and $\pm\frac{2}{3}$. In each of these there is an "odd strand out" - one that is different from the other two. The non-trivial nature of the braid means that there are three distinct positions in the braid which the "odd strand out" can occupy. Hence each of the partially charged states comes in three versions. We will equate these permutations with colour.

- Given these identifications, we can also define generators of the standard model symmetries, $G = SU(3) \times SU(2) \times U(1)$ acting in the standard way on the 30 dimensional space $\mathcal{V}_3^{red}$.

These results suggest identifying $\mathcal{V}_3^L$ with left-handed chiral fermions and $\mathcal{V}_3^R$ with right-handed chiral fermions.

## VII. A POSSIBLE IDENTIFICATION OF THE SECOND GENERATION AND THE VECTOR BOSONS

We have seen that the states which can be imbedded with a minimum number of crossings – i.e. two crossings – give the fermions of the first generation of the standard model. We note that there are exactly the 15 states of the original standard model, with no place for a right-handed neutrino. This is consistent with present data.

It is natural to hypothesize then that the second generation standard model fermions come from the next most complicated states, which have three crossings. These are three stranded braids such as shown in Figure 20. It is straightforward to see that by adding twists to this state one gets a repeat of the pattern for the first generation.

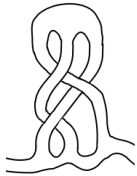

FIG. 20: A possible identification of the muon neutrino as a three crossing state.

The structure of the higher crossing states is under investigation and will be reported elsewhere.

In [1] it is also proposed that the gauge vector bosons of the standard model are composite, and are represented by triplets of ribbons with no crossings. Braids with three ribbons and no crossings are mapped to the bosons of the electroweak interaction. The electroweak interactions between the fermions and the photon and vector bosons are then described by cutting and joining operations on 3-ribbon braids. These preserve the relevant quantum numbers. For more detail, see [1]. How these are incorporated into quantum gravity is presently under study, so we will restrict ourselves here to the remarks below.

## VIII. DISCUSSION

Before closing we want to discuss several issues which bears on the interpretation of these results.

### A. Locality

Locality is a tricky issue in background independent quantum theories of gravity because there is no background metric with which to measure distances or intervals. And it is non-trivial to construct diffeomorphism invariant observables that measure local properties of fields.

This is brought out by the topological nature of the conserved degrees of freedom we have described here. The reader may ask the following question: It is common to encounter such topological degrees of freedom but doesn't their topological nature mean they are of limited interest? And to what extent will they interactions be able to be characterized as local?

The origin of these issues is that there are two notions of locality that may be relevant. In a given graph or ribbon graph there will be a notion of locality: two trinions are neighbors if they are connected, or in a graph two nodes are neighbors if they are connected by a link. We can call this *microlocality.* In the theories we study here, as well as in loop quantum gravity and spin foam models generally, the dynamics is generated by moves that are local in this microscopic sense.

But if this is to be a good theory there should be a notion of classical spacetime geometry that emerges from the quantum geometry. This will give rise to another notion of locality, which we may characterize as *macrolocality.*

The question is then whether there is any guarantee that these two notions of locality will coincide? This indeed seems unlikely, because the quantum states which are expected to represent classical spacetimes are to be constructed from superpositions of graph states each of which carries its own notion of locality.

The noiseless subsystem viewpoint we have used here brings in new physical understanding of this question. The braids are indeed not local with respect to the micro-locality defined by the generators of $\mathcal{A}_{\text{evol}}$. There is, however, no reason that this micro-locality coincides with the macro-locality of the effective spacetime. Instead, the notion of macro-locality should be defined directly from the interactions of the noiseless subsystems that we identify with the elementary particles.

Indeed, it is perhaps misleading to describe $\mathcal{H}$ as a quantum spacetime. It is quantum but also it is pregeometric. The idea is then to not try to regard the quantum geometry as a "bumpy" or "quantum" spacetime. Instead, we simply regard $\{\mathcal{H}, \mathcal{A}_{\text{evol}}\}$ as a quantum information processing system. When we do, a new possibility appears of a dual viewpoint: the locality of the spacetime is to be given and identified by the braid interactions in $\mathcal{H}^B$. It is $\mathcal{A}_{\text{evol}}$ that is non-local with respect to our spacetime!

To see the advantage of this approach we should consider that for one of these models of quantum gravity to have a good low energy limit it must be the case that the emergent symmetries that act on the space of noiseless subsystems includes an approximate translation invariance in space and time. In this case the conserved quantities will include momenta and energy. From the present NS perspective these translation symmetries should emerge as additional symmetries which protect the degrees of freedom we have identified as elementary particles. This will guarantee that the interactions among the particles conserve the emergent notions of energy and momentum. It is this possibility that allows us to use topological conservation numbers to represent matter.

The test of this approach will be whether it can be worked out in detail, so we will not say more here about how translation invariance is to emerge. But we may note that this scenario is very different from the one that has been commonly assumed in many discussions about how space and time are to emerge from background independent models of quantum geometry. It makes no use of the common assumption that the quantum geometry describes a bumpy classical geometry so that the micro-locality of $\mathcal{A}_{\text{evol}}$ ought to coincide with that of the effective spacetime up to Planck scale corrections.

In summary, there are distinct ways for a spacetime geometry to emerge from a quantum theory. At one end of the spectrum lies the expectation that classical spacetime geometry will emerge as the classical/low energy limit of quantum general relativity (as in Loop Quantum Gravity) or a discrete and quantum version of Einstein's theory (as in Causal Dynamical Triangulations [19]). Matter fields are to be added and coupled to the quantum geometry. At the other end, one may expect that the emergent spacetime is the collection of events that are the interactions of the excitations of an underlying pre-spacetime quantum theory, with matter being also emergent as these same excitations. That such an emergent spacetime can be dynamical has recently been investigated in [20, 21].

### B. The mass matrix

In order to arrive at the standard model fermions we have had to make two assumptions about the mass matrix. These should be justified in a fundamental theory. To discuss this we have to notice that the mass matrix can only arise at the level that an effective dynamics in spacetime emerges. As we noted above, this requires that there be in the low energy limit an emergent translation invariance in space and time. This will imply the conservation of energy and momentum for small excitations around the ground state. When the effective Hamiltonian $H$ is evaluated on the states described here at zero momentum it will give us a mass matrix.

The only thing we can say about the mass matrix at this stage is that it should be constructed from topological invariants of the states which are preserved under evolution. Motivated by the spectrum of standard model particles we make the following hypotheses about the mass matrix in this class of theories: (1) Given two braids as just described, which have the same number of strands and twistings, but differ by the number of crossings, the mass will increase with the number of crossings. (2) Non-trivially braided states with both positive and negative twisted strands incident on the same vertex should have a heavy mass $M$. All other states are light relative to the scale $M$.

### C. Emergence of interactions

Another feature of the present proposal that we would like to discuss is the fact that the invariant quantities we find are quantum numbers and not individual identified particles. That is, the in and out states (preserving the relevant quantum numbers) of a Feynman diagram are equivalent. More precisely, the very fact that we are getting $\mathcal{H}^B$ as a noiseless subsystem means that $\mathcal{A}_{\text{evol}}$ commutes with operators on $\mathcal{H}^B$ and hence $\mathcal{A}_{\text{evol}}$ cannot tell us anything about the Standard Model dynamics. In order to describe interactions we will have to weaken the notion of a noiseless subsystem to a more realistic notion of an approximate noiseless subsystem. There are two ways that this could be done. The first is that additional local moves need to be added to $\mathcal{A}_{\text{evol}}$ to give the standard model interactions. The second, more elegant possibility is that the gauge interactions are also emergent and arise at the level that we have a notion of macrolocality that emerges from the interactions of the approximate noiseless subsystems. Both possibilities are under investigation.

### D. The anomaly issue

Finally, we recall that in standard preon models anomaly matching conditions need to be satisfied [12]. This is because the fundamental theory that binds the preons is a conventional, background dependent quantum field theory. In the present setup, the "preon-like" objects appear only at the level of the quantum pregeometry. Individual ribbons do not exist as persistent states and hence will not appear in an effective field theory that describes the dynamics of the persistent states below the Planck scale. So while there can be no anomaly in that effective field theory, we already know there is none, as it is the standard model. As the preons are bound by the dynamics of quantum geometry rather than gauge fields, no issue of anomaly matching arises.

## IX. CONCLUSIONS AND OPEN ISSUES

The results presented here tell us that theories that satisfy the conditions described in section 2 above, may be *already unified*. Most remarkably, with rather mild conditions imposed, the emergent particle degrees of freedom include the first generation fermions and vector bosons of the standard model.

It is worth noting that this result has been achieved through a combination of quite conservative steps. On the other hand, there are many open questions to be resolved if this proposal corresponds to reality. We here list a few.

- How may an effective low energy dynamics emerge? As we discussed in the previous section, if the microscopic dynamics is contained in the evolution algebra, it acts trivially on the sector of the theory that contains braids and twists, meaning that (what we call) the effective theory is completely decoupled from (what we call) the microscopic theory. One would like to know the consequences of a more realistic setup, presumably involving an enlargement of $\mathcal{A}_{\text{evol}}$ so that the sector containing the standard model quantum numbers is approximately noiseless. This case would also give results on the effective dynamics.

- How does the usual quantum statistics arise? This is an important open problem, whose solution may require a further specification of the dynamics so as to satisfy conditions of the topological spin statistics theorems.

- How bound are the quarks and leptons? Does the theory predict the existence of exotic states such as stable fractionally charged particles?

- Is there a composite Higgs or some other mechanism for dynamical symmetry breaking? The states described here naturally are $\mathcal{CP}$ invariant. How may $\mathcal{CP}$ violation enter the theory? Is there a derivation of the mass matrix in terms of topological invariants?

- These results seem to show that the emergent particle-like degrees of freedom arise whether or not representations of a quantum group labels the graphs and even independently of the actual amplitudes for the different evolution moves. How then should these be chosen? Regarding the choice of quantum group, we would favor the minimal choice, which is that there is no group. There are still states that correspond to products of representations of the spacetime $SU(2)_L \oplus SU(2)_R$, but they must be built up by hand from braided triplets of edges. It is then possible that all the quantum numbers, including the geometric labels used in loop quantum gravity can then be regarded as composites of fundamentally topological properties. This gives a very attractive unification in which both the geometry observables and the quantum numbers of the elementary particles are reduced to simple topological properties such as twisting and braiding of edges.

- Alternatively, if the spacetime geometry is to be constructed from the collection of events that are the interactions of the conserved degrees of freedom, one has to show how gravity and the Einstein equations will appear.

**Acknowledgments**

We are very grateful to Stephon Alexander, Olaf Dreyer, Laurent Freidel, Paul Frampton, Louis Kauffman, Jaron Lanier and Shahn Majid for criticism and encouragement during the course of this work. Research at Perimeter Institute is supported in part by the Government of Canada through NSERC and by the Province of Ontario through MEDT.